\documentclass[journal=nalefd,manuscript=letter]{achemso}

\usepackage[version=3]{mhchem} 
\usepackage{graphicx}
\usepackage{amssymb}
\usepackage{color}
\usepackage{amsmath}
\usepackage{bm}

\author{A. A. Mitioglu}
\affiliation{Laboratoire National des Champs Magn\'etiques Intenses, CNRS-EMFL-UJF-UPS-INSA, Grenoble and Toulouse,
France}\altaffiliation{Institute of Applied Physics, Academiei Str. 5, Chisinau, MD-2028, Republic of Moldova}

\author{P. Plochocka}
\affiliation{Laboratoire National des Champs Magn\'etiques Intenses, CNRS-EMFL-UJF-UPS-INSA, Grenoble and Toulouse,
France}\email{paulina.plochocka@lncmi.cnrs.fr}

\author{\'A. Granados del Aguila}
\affiliation{High Field Magnet Laboratory (HFML - EMFL), Institute for Molecules and Materials, Radboud University,
Toernooiveld 7, 6525 ED NIJMEGEN, The Netherlands}

\author{P. C. M. Christianen}
\affiliation{High Field Magnet Laboratory (HFML - EMFL), Institute for Molecules and Materials, Radboud University,
Toernooiveld 7, 6525 ED NIJMEGEN, The Netherlands}

\author{G. Deligeorgis}
\affiliation{FORTH-IESL, Microelectronics Research Group, P.O. Box 1527, 71110 Heraklion, Crete, Greece}

\author{S. Anghel}\affiliation{Institute of Applied Physics, Academiei Str. 5, Chisinau, MD-2028, Republic of Moldova}

\author{L. Kulyuk} \affiliation{Institute of Applied Physics, Academiei Str. 5, Chisinau, MD-2028, Republic of Moldova}

\author{D. K. Maude}\affiliation{Laboratoire National des Champs Magn\'etiques Intenses, CNRS-EMFL-UJF-UPS-INSA, Grenoble and
Toulouse, France}

\title[\texttt{achemso} demonstration]
{Optical investigation of monolayer and bulk tungsten diselenide (WSe$_{2}$) in high magnetic fields}

\begin{document}

\begin{abstract}
Optical spectroscopy in high magnetic fields $B\leq65$~T is used to reveal the very different nature of carriers in
monolayer and bulk transition metal dichalcogenides. In monolayer WSe$_{2}$, the exciton emission shifts linearly with
the magnetic field and exhibits a splitting which originates from the magnetic field induced valley splitting. The
monolayer data can be described using a single particle picture with a Dirac-like Hamiltonian for massive Dirac
fermions, with an additional term to phenomenologically include the valley splitting. In contrast, in bulk WSe$_{2}$
where the inversion symmetry is restored, transmission measurements show a distinctly excitonic behavior with
absorption to the 1s and 2s states. Magnetic field induces a spin splitting together with a small diamagnetic shift and
cyclotron like behavior at high fields, which is best described within the hydrogen model.
\end{abstract}

 \vspace{0.5cm}


Transition metal dichalcogenides (TX$_2$, T=Mo,W,.. and X=S, Se,..) are quasi two dimensional layered materials with
strong ionic-covalent bonding within a layer. A monolayer is composed of a single layer of transition metal embedded
between two atomic layers of chalcogenides atoms in a trigonal prismatic structure. The weak van der Waals interlayer
coupling nevertheless completely modifies the band structure. Bulk crystals are indirect semiconductors with extremely
weak phonon assisted photoluminescence (PL) emission. In monolayer TX$_2$ the inversion symmetry is broken as the two
sublattices are occupied by one T atom and two X$_{2}$ atoms. The band gap is direct and located in two degenerate
valleys $(K\pm)$ at the corner of hexagonal Brillouin zone and the PL emission has a quantum yield which is four orders
of magnitude larger than that of bulk
crystals.~\cite{Albe02,Mak10,Splendiani10,Cheiwchanchamnangij12,Ramasubramaniam12,Zhao13a} In addition, dielectric
confinement, due to the very different dielectric environment outside of the monolayer, enhances the exciton binding
energy beyond the usual factor of four limit for two-dimensional systems.~\cite{Lin2014,Chernikov2014}

\begin{figure}[]
\begin{center}
\includegraphics[width=0.5\columnwidth]{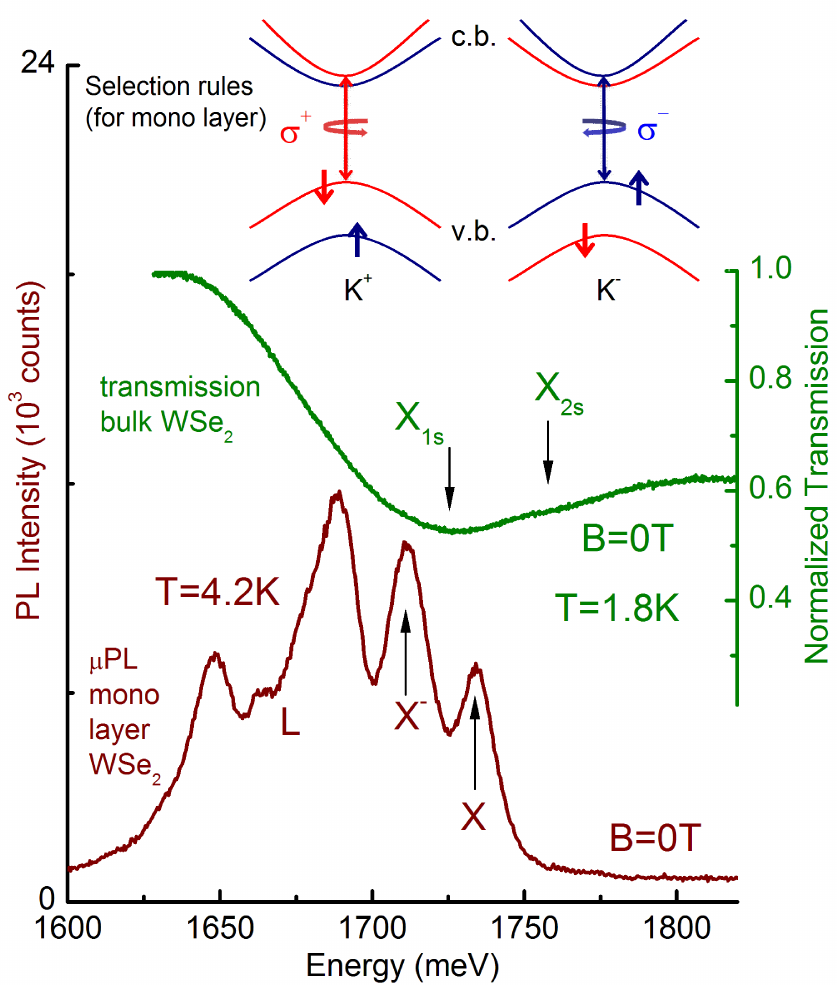}
\end{center}
\caption{Typical low temperature $\mu$PL of monolayer  and transmission spectra of bulk WSe$_{2}$. The inset
illustrates the optical valley selection rules in monolayer WSe$_{2}$ for $\sigma^+$ and $\sigma^-$ circular
polarization of the photons (c.b. labels the conduction band and v.b. labels the valence band).}\label{Fig1}
\end{figure}

Due to the $d$ character of the orbitals, these materials exhibits a strong spin orbit coupling,\cite{Matthesis73}
which when combined with the lack of inversion and time reversal symmetry, breaks the spin degeneracy and leads to a
strong spin orbit induced Zeeman effect.~\cite{Cheiwchanchamnangij12,Zhu11} Hence, the spin and valley degrees of
freedom are coupled which in turn leads to the valley contrasting selection rules for optical interband transitions
with circularly polarized light:~\cite{Xiao12,Wang08} $\sigma^{+}$ couples to the $K+$ valley and $\sigma^{-}$ to
$K^{-}$ respectively (see inset in Fig~\ref{Fig1}). Significant valley polarization was reported recently for monolayer
MoS$_{2}$ and WSe$_{2}$.~\cite{Jones13,Cao12,Mak12} Amongst the transition metal dichalcogenides, WSe$_{2}$ has the
most robust valley polarization, essentially due to the very large spin orbit splitting in the valence band of this
material.~\cite{Jones13}

In transition metal dichalcogenides, the valley degree of freedom can be manipulated and potentially can be used to
replace charge for the transmission of information.~\cite{Wang08,Xiao12,Mak12,Jones13,Xu14,Suzuki14,Behnia2012,Neber12}
For example, the anomalous Hall effect, whose sign depends on the valley index, was recently
demonstrated~\cite{Wu13,Mak14} and an external magnetic field can be used to tune the valley and spin
polarization.~\cite{Aivazian2015} To date, the valley splitting has been investigated for WSe$_2$ and MoSe$_2$ in only
moderately low magnetic fields $B \leq 10$~T.~\cite{Aivazian2015,MacNeill2015,Srivastava15,Li14}

In this letter we compare the optical response of monolayer and bulk WSe$_{2}$ in high magnetic fields up to $B=65$~T.
We show that in monolayer WSe$_{2}$ both the exciton and the trion exhibits a splitting which originates from the
lifting of the valley degeneracy in a magnetic field. The magnetic field has little effect on the valley polarization
of the neutral exciton, whereas it can significantly increase or decrease the valley polarization of the trion
depending on the circular polarization of the excitation light. The linear evolution of the energy of the exciton and
trion features in magnetic field can be described using a Dirac-like Hamiltonian for massive Dirac fermions. In
contrast, in bulk WSe$_{2}$ where the inversion symmetry is restored, we see excitonic behavior with absorption to the
1s and 2s hydrogen like states. Both states show a spin (Zeeman) splitting accompanied by a small diamagnetic shift at
low magnetic fields before shifting as the cyclotron energy in the high field limit.

For the measurements, single layer flakes of tungsten diselenide (WSe$_{2}$) have been obtained by mechanical
exfoliation of bulk 2H-WSe$_2$ (the hexagonal 2H-polytype of tungsten diselenide) single crystals grown using chemical
vapor transport with Bromine as the transport agent. Samples obtained in this way are naturally n-type.~\cite{Evans77}
Two types of experiments have been performed; (i) micro-photoluminescence ($\mu$PL) on direct gap monolayer WSe$_{2}$
in steady state magnetic field up to $30$~T and (ii) macro transmission of thin crystals of indirect gap bulk WSe$_{2}$
in the pulsed magnetic field up to $65$ T. Combining PL and transmission allows us to probe the A-exciton of an
electron-hole pair at the K-points of the Brillouin zone in both monolayer and in bulk WSe$_{2}$ for which PL is
dominated by the indirect gap.

For $\mu$PL measurements the WSe$_2$ monolayer flake was placed in a system composed of piezoelectric $x-y-z$
translation stages and a microscope objective. The $\mu$PL system was cooled in exchange gas to a temperature of
T=$4.2$~K in a cryostat filled with liquid helium which was placed in a resistive magnet producing magnetic fields up
to $B=30$~T. The magnetic field was applied in the Faraday configuration and the sample was illuminated by a laser at
$640$ nm. The excitation power was kept low (of the order of a few hundred of nW) to avoid heating effects. Both the
exciting and collected light were transmitted through a non polarizing cube beam splitter (50:50) placed on the optical
axis of the objective. With this setup the circular polarization of the excitation and detection can be controlled
independently. The diameter of the excitation beam on the sample was of the order of 1~$\mu$m.

The magneto-transmission measurements were performed on thin bulk samples (also obtained by limited mechanical
exfoliation) in pulsed fields $\leq 65$~T ($\simeq 400$~ms). A tungsten halogen lamp provides a broad spectrum in the
visible and near infra-red range and the absorption is measured in the Faraday configuration with the $c$-axis of the
sample parallel to magnetic field. Typical size of the spot was of the order of 200~$\mu$m and the polarization optics
was introduced in situ. The sample was immersed in a helium bath cryostat pumped to a temperature of $\simeq1.8$~K to
avoid noise due to the formation of helium gas bubbles. Note, that we are unable to perform macro transmission
measurements on monolayer WSe$_2$ as only a tiny fraction of the transmitted light would pass through the micron size
flake and our $\mu$PL setup is not adapted for micro-transmission measurements.

For both measurements the collected light was dispersed in a spectrometer equipped with a multichannel CCD camera
cooled to liquid nitrogen temperatures.

Fig~\ref{Fig1} shows a comparison between the emission spectra measured on a single layer flake and the transmission
through the thin bulk crystals of the WSe$_{2}$ all measured at zero magnetic field. In both spectra we observe a
strong resonance around 1730 meV which corresponds well to the energy 1s ($n=1$) state of the A-exciton, previously
reported for monolayer~\cite{Zeng13,Zhao13,Jones13,Urbaszek14} and bulk WSe$_{2}$~\cite{Beal76,Zeng13}. The slightly
higher energy of the A exciton emission observed in monolayer layer WSe$_{2}$ (compared to bulk transmission) is also
in agreement with previous studies~\cite{Zeng13,Zhao13}. We also see the 2s ($n=2$) exciton level in transmission,
previously reported in reflectance measurements.~\cite{Beal76} The separation of the 1s and 2s transitions, equal to
$3R_y^*/4$ in the 3D hydrogen model ($E_n = R_y^*/n^2, n=1,2,..$) provides a rough estimate for the exciton binding
energy in bulk WSe$_2$ of $R_y^*\approx 44$~meV. The monolayer emission spectrum has additional emission lines on the
low energy side. The line around 30 meV lower than the exciton corresponds well to the negatively charged exciton
(trion) emission, as our samples are naturally n-doped.~\cite{Evans77} Trion emission is routinely observed in mono
layers dichalcogenides with an excess concentration of electrons and can be unambiguously identified by its
characteristic power dependence.~\cite{Mak13,Mitioglu13} Moreover, the observed binding energy of the trion corresponds
well to the experimental value previously reported for WSe$_{2}$.~\cite{Jones13,Urbaszek14,Zhu14} Below the trion
emission several peaks are observed which we label as ``L'' as they can be assigned to localized (bound exciton) states
previously reported in WSe$_{2}$ and MoS$_{2}$\cite{Urbaszek14,Zhu14,Mak12}.

\begin{figure}[]
\begin{center}
\includegraphics[width=0.5\columnwidth]{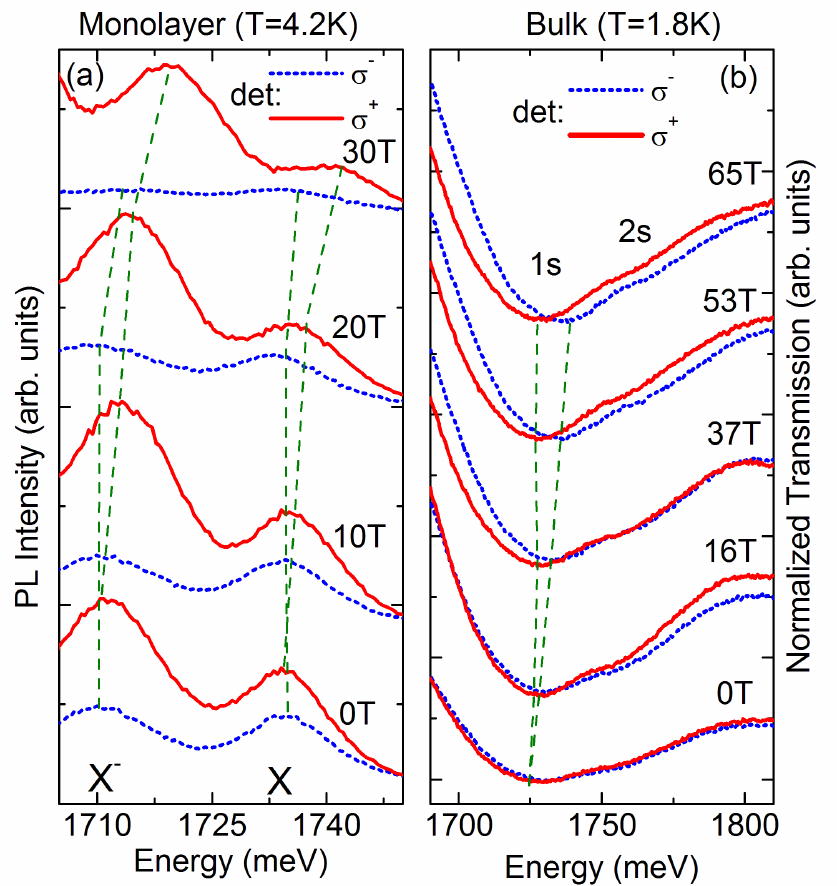}
\end{center}
\caption{Typical low temperature spectra at various magnetic fields of (a) $\mu$PL from monolayer WSe$_{2}$ and (b)
transmission through the thin WSe$_{2}$ bulk crystal, using $\sigma^{+}$ and $\sigma^{-}$ detection. For the PL shown
here $\sigma^{+}$ polarized light was used for excitation while the white light for the transmission measurements was
unpolarized. The dashed lines are drawn as a guide for eye showing the evolution of exciton features with magnetic
field. The spectra have been shifted vertically for clarity.}\label{Fig2}
\end{figure}

\begin{figure}[]
\begin{center}
\includegraphics[width=0.9\columnwidth]{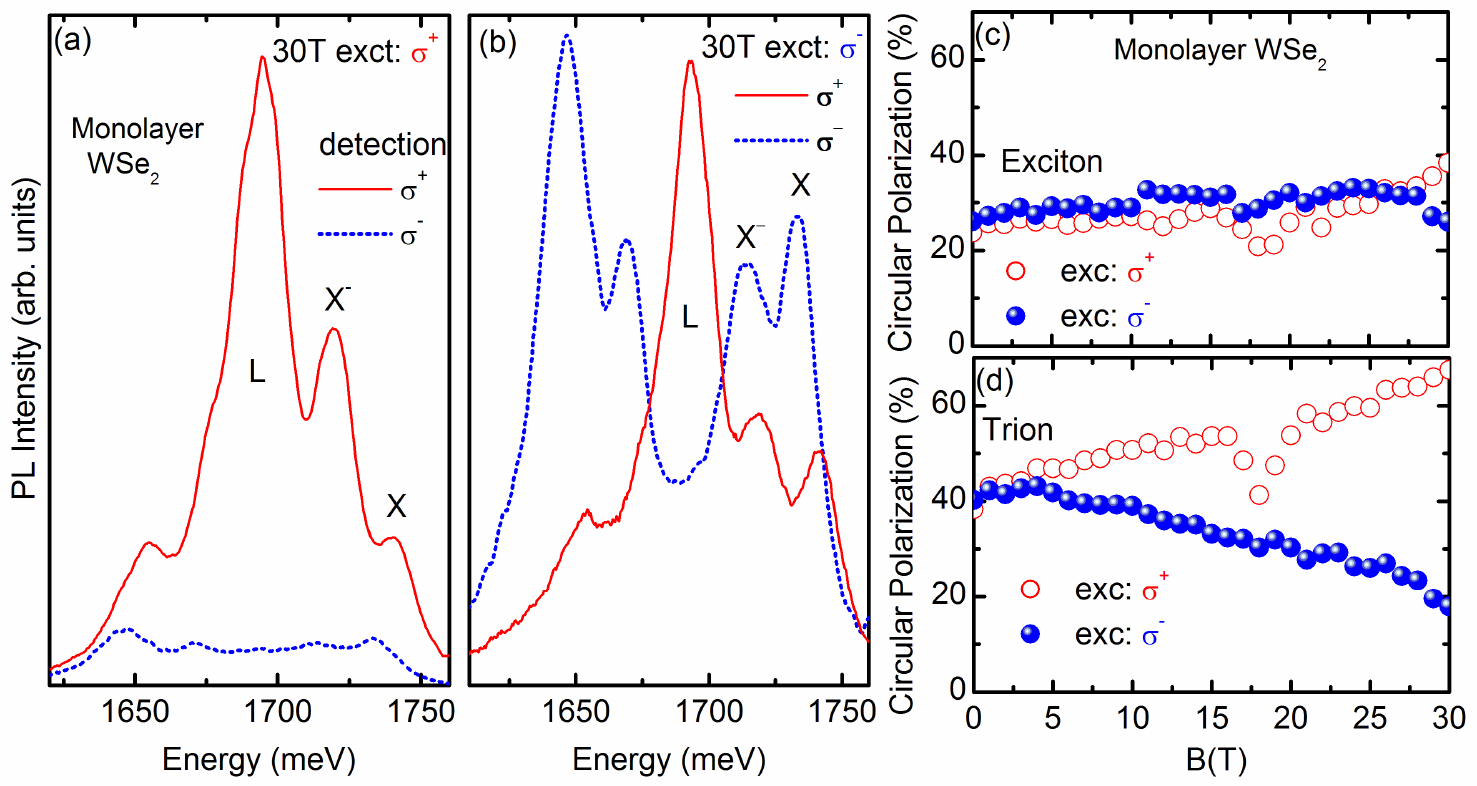}
\end{center}
\caption{(a),(b) Typical $\mu$PL spectra excited $\sigma^{+}$ and $\sigma^{-}$ polarized laser respectively. The
red/blue lines denote the detection in $\sigma^{+}$ and $\sigma^{-}$ respectively. (c),(d) degree of circular
polarization of the neutral and charged exciton respectively. The open symbols denote $\sigma^{+}$ excitation as closed
one correspond to the $\sigma^{-}$ polarization of the excitation.}\label{Fig3}
\end{figure}

In the following we focus on the magnetic field dependence of the exciton transitions. Typical low temperature $\mu$PL
spectra of monolayer WSe$_{2}$ measured in \emph{dc} magnetic fields up to 30~T are presented in Fig~\ref{Fig2}(a). The
results of the transmission measurements through the thin layers of bulk WSe$_{2}$ measured in pulsed magnetic fields
up to 65~T are presented in Fig~\ref{Fig2}(b). For both data sets the solid and dotted lines correspond to $\sigma^{+}$
and $\sigma^{-}$ detection respectively. The photoluminescence was excited using $\sigma^{+}$ polarization while for
the transmission unpolarized white light was used for the excitation.

In both bulk and monolayer crystals the polarization resolved measurements reveal a splitting of the exciton
transition, nevertheless with very different physical origins due to the markedly different optical selection rules in
bulk and monolayer WSe$_2$. Additionally, the monolayer emission shows that, already at zero magnetic field, both the
neutral and charged exciton populations are partially valley polarized as the emission intensity is not the same for
$\sigma^{+}$ and $\sigma^{-}$ detection. This is not surprising, as our circular excitation was relatively close to the
exciton resonance, as also previously reported for WSe$_{2}$.~\cite{Jones13}

In monolayer WSe$_{2}$ the circular polarization $\sigma^\pm$ of the excitation allows us to selectively pump the $K^+$
or $K^-$ valley. The circular polarization of the emission then depends on the inter valley scattering rate and the
radiative recombination lifetime. In the absence of inter valley scattering the emission would be 100\% polarized. The
degree of circular polarization $\textit{P} $ for the exciton and trion emission can be defined as
\begin{equation}
 \textit{P} = \frac{I_{\sigma^{+}}-I_{\sigma^{-}}}{I_{\sigma^{+}} + I_{\sigma^{-}}},\label{EqPolarisation}
\end{equation}
where $I_{\sigma^{\pm}}$ is the intensity of the photoluminescence for a given circular (detection) polarization
obtained by fitting the spectra with a Gaussian function for each value of magnetic field.

We have analyzed data for all four possible combinations of the circular polarization of the excitation and detection.
Typical polarization resolved $\mu$PL spectra measured at 30~T for $\sigma^{+}$ and $\sigma^{-}$ are presented in
Fig~\ref{Fig3}(a) and (b) respectively. Note that the direction of the magnetic field was the same for all
measurements. The required polarization combination was selected by rotating the excitation and detection polarization
optics located outside of the cryostat. As in zero field, the 30~T raw data clearly shows that both neutral and charge
exciton exhibits a significant degree of circular polarization.

The full magnetic field dependence of calculated degree of circular
polarization of the neutral and charged exciton calculated using
Equation $1$ is presented in Fig~\ref{Fig3}(c),(d) respectively.
Both neutral and charged exciton shows a valley polarization in zero
magnetic field. For the exciton $ \textit{P} = 25\%$ and for the
charged exciton $ \textit{P} = 40\%$. Although we are not exciting
exactly at the resonance it has been shown that WSe$_{2}$, which has
the largest spin orbit induced spin splitting in both valence and
the conduction bands of any member of the TX$_{2}$ family, has a
valley polarization which is the most robust of all the transition
metal dichalcogenides.~\cite{Jones13} For the exciton the degree of
the circular polarization remains almost unchanged with the magnetic
field suggesting that magnetic field has little influence on the
inter valley exciton scattering rate. On the other hand the degree
of polarization is trion is strongly influenced with the
polarization of the $\sigma^+$ excited emission (pumping K+ valley)
increasing strongly, accompanied by an equally strong reduction in
the degree of polarization for $\sigma^-$ excitation (pumping K-
valley). This suggests, that although the magnetic field does not
modify the inter valley exciton scattering rate, it can nevertheless
modify the probability for trion formation, most likely via its
influence on the excess electron population of the spin orbit split
spin levels in the conduction band for each valley.

The evolution of the exciton and trion polarization reported here is somewhat different from previous investigations of
WSe$_2$~\cite{Aivazian2015} and MoSe$_2$~\cite{MacNeill2015} which might be due to the low magnetic fields employed ($B
\leq 7$~T). For example, in our data, the true behavior of the trion polarization is only revealed for magnetic fields
$B \geq 5$~T. Additionally, the polarization depends sensitively upon the exact excitation used, together with sample
dependent parameters such as the valley scattering rate.

\begin{figure}[]
\begin{center}
\includegraphics[width=0.5\columnwidth]{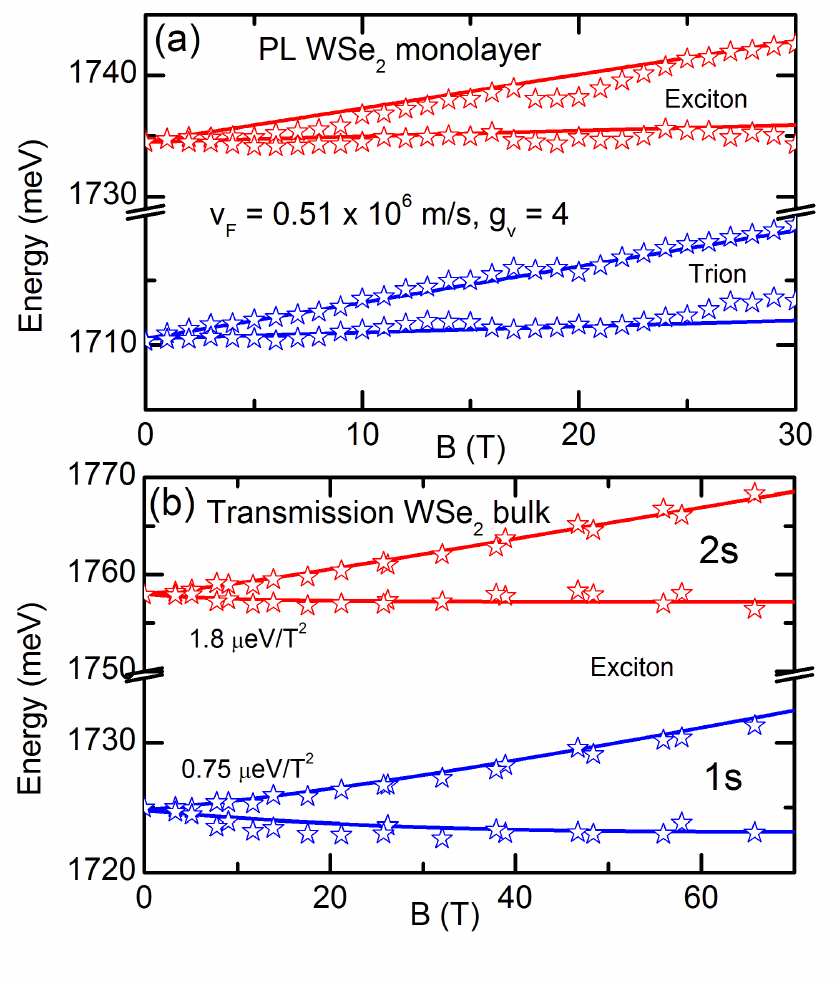}
\end{center}
\caption{(a) Magnetic field dependence of the energy of the
polarization resolved exciton emission in monolayer WSe$_2$ measured
at $T=4.2$~K. The solid lines are fits to the data using the
Hamiltonian for massive Dirac fermions. (b) Magnetic field
dependence of transmission in bulk WSe$_2$ measured at $T=1.8$~K.
The solid lines are fits to the data using Equation $3$ to model the
diamagnetic (quadratic) shift at low magnetic fields followed by a
linear Landau like behavior at high magnetic fields.}\label{Fig4}
\end{figure}

The energy of the polarization resolved exciton and trion emission for monolayer WSe$_2$ is shown in
figure~\ref{Fig4}(a). Both the exciton and trion energy shifts linearly to higher energy with increasing magnetic field
showing a pronounced splitting which also evolves linearly with magnetic field. In the transition metal dichalcogenides
the carriers behave as massive Dirac fermions which can be described by a Dirac like
Hamiltonian.~\cite{Xiao12,Rose2013} In this simple picture the magnetic field quantizes the energy of the carriers in
the conduction and valence bands into Landau levels with energy,
\begin{equation}
E_{\lambda}^{\pm K} = \lambda \sqrt{\Delta^2 + n \varepsilon^2} \pm
\frac{1}{2} g_v \mu_B B,\label{EqELL}
\end{equation}
where $\lambda=\pm 1$ designates the conduction/valence band, $\Delta$ is half of the band gap and
$\varepsilon=\sqrt{2} \hbar v_F/\ell_B$ is the characteristic magnetic energy. Here $n$ is the Landau level index,
$v_F$ is the Fermi velocity and $\ell_B = (\hbar/e B)^{1/2}$ is the magnetic length. Note in this single particle
picture $2\Delta$ is obtained from the observed zero magnetic field transition energy to which the exciton binding
energy in monolayer WSe$_2$ of $E_B=370$~meV has to be added~\cite{He2014}.

\begin{figure}[]
\begin{center}
\includegraphics[width=0.5\columnwidth]{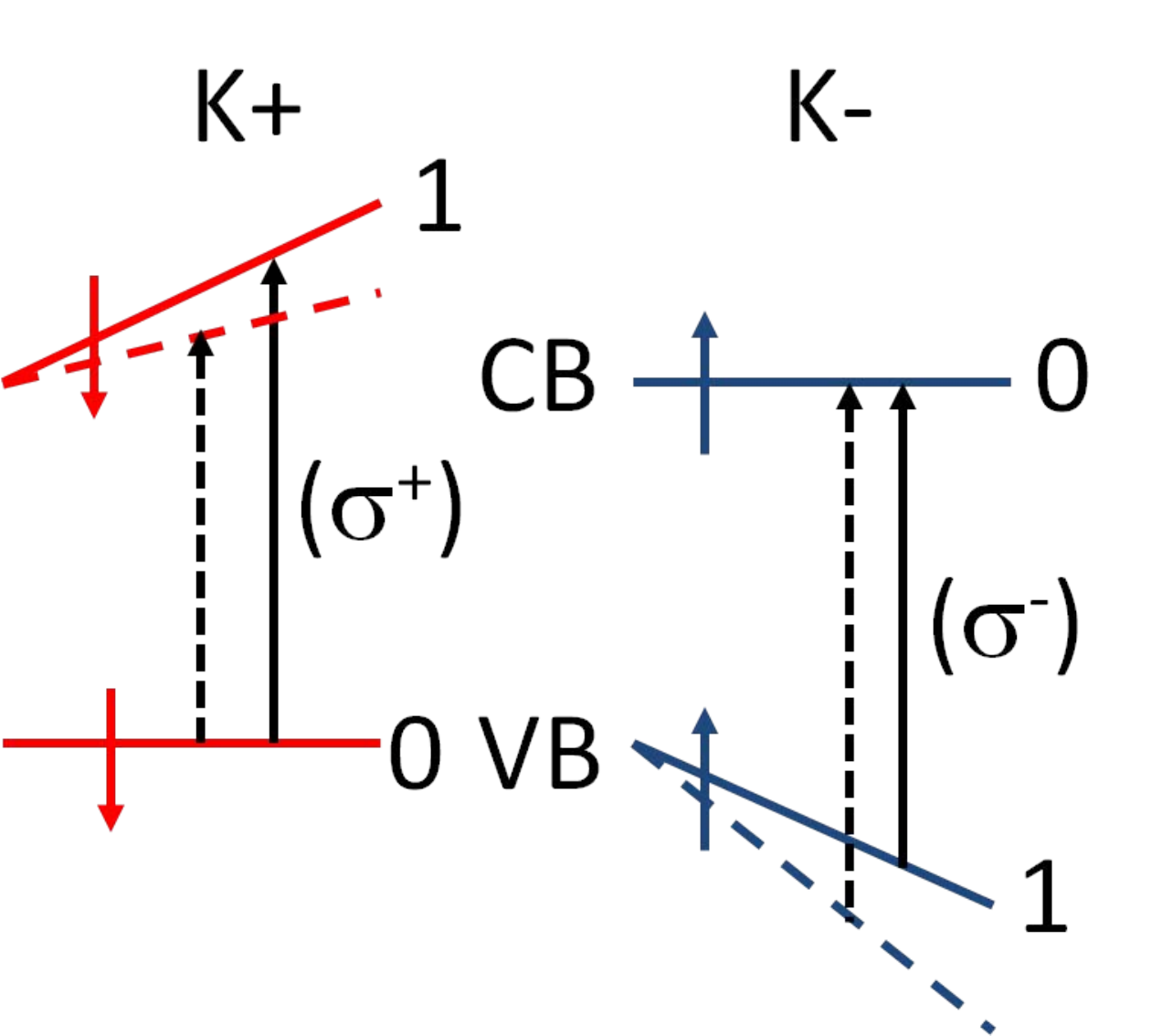}
\end{center}
\caption{Schematic showing evolution of the $n=0,1$ Landau levels of massive Dirac fermions in a magnetic field
together with the dipole allowed $0 \rightarrow 1$ and $1 \rightarrow 0$ transitions which select the valley involved
(solid lines are Landau levels, solid arrows indicate the transitions). Spin is conserved so that the Zeeman energy
(not shown here) does not split the transition. Any asymmetry of the electron/hole Dirac cones, which modifies the
energy of the $n=1$ Landau levels, would split the transition as indicated by the dashed lines/arrows.}\label{Fig5}
\end{figure}

As in graphene, in the transition metal dichalcogenides the selections rules for dipole allowed optical transitions are
$\Delta n = \pm 1$, \emph{i.e.} the Landau level index changes by one. In this case spin is conserved and any spin
splitting of the Landau levels will not give rise to a splitting of the optical transition unless the electron and hole
g-factors are significantly different. Thus, the observed splitting of the transitions is included simply as a
phenomenological valley splitting $\pm \frac{1}{2} g_v \mu_B B$ where $g_v$ is the effective valley g-factor. The $n>0$
Landau levels always occur in pairs with one in each valley in both the conduction and valence bands. The $n=0$ level
is special and there is a single $n=0$ Landau level per valley. For both spin levels, the $n=0$ Landau level is fixed
at the top of the valence band for the $+K$ valley and at the bottom of the conduction band for the $-K$
valley.\cite{Rose2013} Thus, in a magnetic field resonant $\sigma^\pm$ excitation selects $0 \rightarrow 1$ or $1
\rightarrow 0$ optical transitions and therefore selecting the valley as in the zero magnetic field case (see schematic
in Figure~\ref{Fig5}).

The solid lines in figure~\ref{Fig4}(a) are the transitions energies
calculated from the energies of the $0 \rightarrow 1$ and $1
\rightarrow 0$ transitions from Equation $2$ with a Fermi velocity
$v_F = (0.51 \pm 0.02) \times 10^6$~m/s and an effective valley
g-factor $g_v =4 \pm 0.5$. We have included excitonic effects simply
by subtracting the exciton (trion) binding energy from the
calculated transition energy. The parameter set provides a
reasonable fit to both the exciton and trion emission data. The
transition energies evolve linearly with magnetic field which can be
understood as follows: The band gap of WSe$_2$ is large so that for
the magnetic fields of interest here $\Delta^2 \gg \varepsilon^2$ so
that the orbital contribution to the Landau level energy is well
approximated by a Taylor expansion
\begin{equation}
 E_n = \pm \Delta(1 + n \varepsilon^2/\Delta^2)^{\frac{1}{2}} \simeq \pm (\Delta + n \hbar e B v_F^2/\Delta).\nonumber
\end{equation}
Thus, the carriers behave as massive Dirac fermions with an
effective mass $m^* = \Delta/v_F^2 \simeq 0.7 m_e$ where $m_e$ is
the free electron mass. Note, that this result can also be obtained
by comparison with the Klein-Gordon equation $E^2 = m^2 c^4 + p^2
c^2$ replacing the speed of light with the Fermi velocity. For this
reason $\Delta \equiv m c^2$ is often referred to as the ``mass
gap''.

The observed splitting is asymmetric due to the orbital (Landau level) contribution which is always positive. The low
energy peak almost does not move in magnetic field as the orbital and valley contributions are of similar size but of
opposite sign. Li et al.~\cite{Li14} observe a similar valley splitting in MoSe$_2$ of 0.24meV/T but with a symmetric
splitting in fields up to 10T. This may suggest that the orbital contribution is quenched in their measurements, either
due to the higher temperature used ($T=10$~K) or increased disorder in their sample ($\omega_c \tau < 1$).

In the literature the splitting of the transition is often described, using slightly different language, in terms of a
two band tight binding model in which the atomic $d$-orbitals which form the valence band have an in plane angular
momentum $m\hbar$ with $m= \pm 2$ according to the valley index. In magnetic field, the valley splitting is then $4
\mu_B B$ which agrees within error with the splitting observed here (we estimate from the fit that $g_v = 4.0 \pm
0.5$). From low field measurements ($B<10T$) Srivastava and co-workers find a similar value for the valley splitting of
WSe$_2$ ($g_v = 4.3$),~\cite{Srivastava15} while a factor of two lower value ($\simeq2\mu_B B$) has been the published
recently by Aivazian and co-workers.~\cite{Aivazian2015} In the tight binding model, the valley magnetic moment due to
the self rotation of the wave packet gives no contribution to the splitting provided the electron and hole masses are
identical. Aivazian \emph{et al.}~\cite{Aivazian2015} invoked different electron and hole masses to correctly predict
the observed splitting in their samples. Similarly, in our Hamiltonian any asymmetry of the electron and hole Dirac
cones would lead to a splitting of the $0 \rightarrow 1$ and $1 \rightarrow 0$ transitions (see Figure~\ref{Fig5}).
However, as our data corresponds within error to the expected splitting of $4 \mu_B B$, this suggest that electron-hole
asymmetry in WSe$_2$ is small and makes no significant contribution to the observed valley splitting.

For bulk WSe$_2$, the dipole selection rules are $\Delta n=0$ so that the angular momentum of the photon is used to
flip the spin. The energy of the polarization resolved exciton features in transmission for bulk WSe$_2$ shown in
figure~\ref{Fig4}(b) clearly shows signs of excitonic effects with quadratic diamagnetic shift at low magnetic fields
which becomes more linear at high fields. The evolution of the data in magnetic field can be well described using
\begin{equation}
\Delta E_{\uparrow \downarrow} = -\frac{\hbar \omega_0}{2} + \frac{\hbar}{2} \sqrt{\omega_0^2 + \omega_c^2} \pm
\frac{1}{2} g_s \mu_B B,\label{Eq:FD}
\end{equation}
where $\hbar \omega_0$ is an energy which controls the diamagnetic shift, $\hbar \omega_c = e B /m^*$ is the cyclotron
energy and $m^*$ is the reduced exciton mass. The observed spin splitting is included via an effective g-factor $g_s$.

Neglecting spin for the moment, at high magnetic field $\omega_c^2 \gg \omega_0^2$ and the energy is well approximated
by $\hbar \omega_c/2$. At low fields when $\omega_0^2 \gg \omega_c^2$ to first order a Taylor expansion gives the
diamagnetic shift
\begin{equation}
\Delta E \simeq \frac{\hbar e^2 B^2}{4 \omega_0 m^{*2}} = e^2 a_0^2
B^2.\nonumber
\end{equation}
Here we have defined a length $a_0 = \sqrt{\hbar / 4m^{*2}
\omega_0}$ which can be identified with the effective Bohr radius in
the hydrogen model.

The solid lines in figure~\ref{Fig4}(b) are the transition energies
calculated using Equation $3$. For the 1s absorption, we use an
effective mass $m^* = 0.7 m_e$ as in monolayer WSe$_2$. The spin
splitting can be reproduced using an effective g-factor $g_s = 2.3$
and the diamagnetic shift reproduced using $\hbar \omega_0 = 8$~meV.
This gives an estimation of the diamagnetic shift at low magnetic
fields of $\Delta E \approx 0.9\mu$V/T$^2$. The effective Bohr
radius is $a_0 \approx 1.8$~nm from which we can obtain a rough
estimate of the exciton binding energy in the hydrogen model $e^2/8
\pi \epsilon_{r} \epsilon_{0} a_{0} \approx 56$~meV using a
dielectric constant $\epsilon_r \simeq 7$ for bulk 2H-WSe2. This
value agrees well with the exciton binding energy of $\simeq 55$~meV
determined directly from the $n=1,2,3$ hydrogen series of levels in
reflectance measurements on bulk 2H-WSe$_2$ crystals.\cite{Beal76} A
similar binding energy ($\simeq 56 meV$) has been
reported~\cite{Goto2000} in high magnetic field measurements up to
150~T on bulk MoS$_2$, however with a much smaller diamagnetic shift
$\simeq 0.2\mu$eV/T$^2$. The approximately four fold smaller
diamagnetic shift is however, fully consistent with our data on
WSe$_{2}$ when the smaller Bohr radius (1.28~nm) and smaller reduced
exciton mass ($0.4 m_e$) in MoS$_2$ are taken into
account~\cite{Goto2000}.

For completeness, we have also fitted the 2s data using the same effective mass $m^* = 0.7 m_e$, a slightly larger
effective g-factor is required $g_s = 2.8$, with a smaller value of $\hbar \omega_0 = 2$~meV. The estimated diamagnetic
shift $\Delta E \approx 3.4\mu$V/T$^2$ is larger due to the more delocalized nature of the 2s wave function which
extends over $\approx 3.7$~nm.

In conclusion, we have investigated monolayer and bulk WSe$_{2}$ in high magnetic fields up to $B=65$~T using optical
spectroscopy. In monolayer WSe$_{2}$, the exciton emission exhibits a splitting which originates from lifting of the
valley degeneracy by the magnetic field. The linear evolution of the energy of the exciton features in magnetic field
can be described using a Dirac-like Hamiltonian for massive Dirac fermions with a Fermi velocity of $0.51 \times
10^6$m/s and an effective valley $g$-factor $g_v=4$ for an assumed effective spin $s=1/2$ system. The measured Fermi
velocity can be used to estimate the effective hopping integral $t$ of the tight binding Hamiltonian using $at=\hbar
v_F$ where $a=3.31{\AA}$. The measured value of $\hbar v_F =3.36{\AA}$eV suggests $t=1.02$~eV which is roughly 15\%
less than the effective hopping integral predicted from first principle band structure
calculations~\cite{Xiao12,Rose2013}. In contrast, in bulk WSe$_{2}$ where the inversion symmetry is restored,
transmission measurements show that the exciton exhibits a spin (Zeeman) splitting and the exciton 1s and 2s features
show a small diamagnetic shift which allows us to determine the exciton binding energy of around 56~meV within the
three dimensional hydrogen model. In two dimensions, the binding energy is enhanced by at most a factor of four, so the
predicted 2D exciton binding energy ($4 \times 56 \approx 224$~meV) is significantly less than the established value of
370~meV in monolayer WSe$_2$.~\cite{Beal76} This suggests that dielectric screening (image charge), due to the very
different dielectric environment outside of the material,~\cite{Lin2014,Chernikov2014} significantly enhances the
exciton binding energy in monolayer transition metal dichalcogenides.

\acknowledgement

This work was partially supported by  Programme Investissements d'Avenir under the program ANR-11-IDEX-0002-02 -
reference ANR-10-LABX-0037-NEXT, ANR JCJC project milliPICS, the Region Midi-Pyr\'en\'ees under contract MESR 13053031
, and STCU project 5809, We also acknowledge the support of HFML-RU/FOM, member of the European Magnetic Field
Laboratory (EMFL) and support by EuroMagNET II under the EU Contract Number 228043. One of us (A.A.M.) was partially
supported during his visit to HMFL to perform these measurements by a ``Bourse d'excellence EOLE du R\'eseau
franco-n\'eerlandais''. Finally, it is our pleasure to thank S. George for her technical assistance with the pulsed
field transmission measurements.

\suppinfo{Four figures in which Gaussians are fitted to the raw data to extract the energy of the exciton and trions
features are available as supplementary information.}


\begin{mcitethebibliography}{37}
\providecommand*\natexlab[1]{#1} \providecommand*\mciteSetBstSublistMode[1]{}
\providecommand*\mciteSetBstMaxWidthForm[2]{} \providecommand*\mciteBstWouldAddEndPuncttrue
  {\def\EndOfBibitem{\unskip.}}
\providecommand*\mciteBstWouldAddEndPunctfalse
  {\let\EndOfBibitem\relax}
\providecommand*\mciteSetBstMidEndSepPunct[3]{} \providecommand*\mciteSetBstSublistLabelBeginEnd[3]{}
\providecommand*\EndOfBibitem{} \mciteSetBstSublistMode{f}
\mciteSetBstMaxWidthForm{subitem}{(\alph{mcitesubitemcount})} \mciteSetBstSublistLabelBeginEnd
  {\mcitemaxwidthsubitemform\space}
  {\relax}
  {\relax}

\bibitem[Albe and Klein(2002)Albe, and Klein]{Albe02}
Albe,~K.; Klein,~A. \emph{Phys. Rev. B} \textbf{2002}, \emph{66}, 073413\relax \mciteBstWouldAddEndPuncttrue
\mciteSetBstMidEndSepPunct{\mcitedefaultmidpunct} {\mcitedefaultendpunct}{\mcitedefaultseppunct}\relax \EndOfBibitem
\bibitem[Mak \latin{et~al.}(2010)Mak, Lee, Hone, Shan, and Heinz]{Mak10}
Mak,~K.~F.; Lee,~C.; Hone,~J.; Shan,~J.; Heinz,~T.~F. \emph{Phys. Rev. Lett.}
  \textbf{2010}, \emph{105}, 136805\relax
\mciteBstWouldAddEndPuncttrue \mciteSetBstMidEndSepPunct{\mcitedefaultmidpunct}
{\mcitedefaultendpunct}{\mcitedefaultseppunct}\relax \EndOfBibitem
\bibitem[Splendiani \latin{et~al.}(2010)Splendiani, Sun, Zhang, Li, Kim, Chim,
  Galli, and Wang]{Splendiani10}
Splendiani,~A.; Sun,~L.; Zhang,~Y.; Li,~T.; Kim,~J.; Chim,~C.-Y.; Galli,~G.;
  Wang,~F. \emph{Nano Letters} \textbf{2010}, \emph{10}, 1271--1275\relax
\mciteBstWouldAddEndPuncttrue \mciteSetBstMidEndSepPunct{\mcitedefaultmidpunct}
{\mcitedefaultendpunct}{\mcitedefaultseppunct}\relax \EndOfBibitem
\bibitem[Cheiwchanchamnangij and Lambrecht(2012)Cheiwchanchamnangij, and
  Lambrecht]{Cheiwchanchamnangij12}
Cheiwchanchamnangij,~T.; Lambrecht,~W.~R. \emph{Phys. Rev. B} \textbf{2012},
  \emph{85}, 205302\relax
\mciteBstWouldAddEndPuncttrue \mciteSetBstMidEndSepPunct{\mcitedefaultmidpunct}
{\mcitedefaultendpunct}{\mcitedefaultseppunct}\relax \EndOfBibitem
\bibitem[Ramasubramaniam(2012)]{Ramasubramaniam12}
Ramasubramaniam,~A. \emph{Phys. Rev. B} \textbf{2012}, \emph{86}, 115409\relax \mciteBstWouldAddEndPuncttrue
\mciteSetBstMidEndSepPunct{\mcitedefaultmidpunct} {\mcitedefaultendpunct}{\mcitedefaultseppunct}\relax \EndOfBibitem
\bibitem[Zhao \latin{et~al.}(2013)Zhao, Ribeiro, Toh, Carvalho, Kloc,
  Castro~Neto, and Eda]{Zhao13a}
Zhao,~W.; Ribeiro,~R.~M.; Toh,~M.; Carvalho,~A.; Kloc,~C.; Castro~Neto,~A.~H.;
  Eda,~G. \emph{Nano Letters} \textbf{2013}, \emph{13}, 5627--5634\relax
\mciteBstWouldAddEndPuncttrue \mciteSetBstMidEndSepPunct{\mcitedefaultmidpunct}
{\mcitedefaultendpunct}{\mcitedefaultseppunct}\relax \EndOfBibitem
\bibitem[Lin \latin{et~al.}(2014)Lin, Ling, Yu, Huang, Hsu, Lee, Kong,
  Dresselhaus, and Palacios]{Lin2014}
Lin,~Y.; Ling,~X.; Yu,~L.; Huang,~S.; Hsu,~A.~L.; Lee,~Y.-H.; Kong,~J.;
  Dresselhaus,~M.~S.; Palacios,~T. \emph{Nano Letters} \textbf{2014},
  \emph{14}, 5569--5576\relax
\mciteBstWouldAddEndPuncttrue \mciteSetBstMidEndSepPunct{\mcitedefaultmidpunct}
{\mcitedefaultendpunct}{\mcitedefaultseppunct}\relax \EndOfBibitem
\bibitem[Chernikov \latin{et~al.}(2014)Chernikov, Berkelbach, Hill, Rigosi, Li,
  Aslan, Reichman, Hybertsen, and Heinz]{Chernikov2014}
Chernikov,~A.; Berkelbach,~T.~C.; Hill,~H.~M.; Rigosi,~A.; Li,~Y.;
  Aslan,~O.~B.; Reichman,~D.~R.; Hybertsen,~M.~S.; Heinz,~T.~F. \emph{Phys.
  Rev. Lett.} \textbf{2014}, \emph{113}, 076802\relax
\mciteBstWouldAddEndPuncttrue \mciteSetBstMidEndSepPunct{\mcitedefaultmidpunct}
{\mcitedefaultendpunct}{\mcitedefaultseppunct}\relax \EndOfBibitem
\bibitem[Mattheiss(1973)]{Matthesis73}
Mattheiss,~L. \emph{Phys. Rev. B} \textbf{1973}, \emph{8}, 3719--3740\relax \mciteBstWouldAddEndPuncttrue
\mciteSetBstMidEndSepPunct{\mcitedefaultmidpunct} {\mcitedefaultendpunct}{\mcitedefaultseppunct}\relax \EndOfBibitem
\bibitem[Zhu \latin{et~al.}(2011)Zhu, Cheng, and Schwingenschl\"ogl]{Zhu11}
Zhu,~Z.; Cheng,~Y.; Schwingenschl\"ogl,~U. \emph{Phys. Rev. B} \textbf{2011},
  \emph{84}, 153402\relax
\mciteBstWouldAddEndPuncttrue \mciteSetBstMidEndSepPunct{\mcitedefaultmidpunct}
{\mcitedefaultendpunct}{\mcitedefaultseppunct}\relax \EndOfBibitem
\bibitem[Xiao \latin{et~al.}(2012)Xiao, Liu, Feng, Xu, and Yao]{Xiao12}
Xiao,~D.; Liu,~G.-B.; Feng,~W.; Xu,~X.; Yao,~W. \emph{Phys. Rev. Lett.}
  \textbf{2012}, \emph{108}, 196802\relax
\mciteBstWouldAddEndPuncttrue \mciteSetBstMidEndSepPunct{\mcitedefaultmidpunct}
{\mcitedefaultendpunct}{\mcitedefaultseppunct}\relax \EndOfBibitem
\bibitem[Yao \latin{et~al.}(2008)Yao, Xiao, and Niu]{Wang08}
Yao,~W.; Xiao,~D.; Niu,~Q. \emph{Phys. Rev. B} \textbf{2008}, \emph{77},
  235406\relax
\mciteBstWouldAddEndPuncttrue \mciteSetBstMidEndSepPunct{\mcitedefaultmidpunct}
{\mcitedefaultendpunct}{\mcitedefaultseppunct}\relax \EndOfBibitem
\bibitem[Jones \latin{et~al.}(2013)Jones, Yu, Wu, Aivazian, Ross, Zhao, Yan,
  Mandrus, Xiao, Yao, and Xu]{Jones13}
Jones,~A.~M.; Yu,~N.~J.,~Hongyi amd~Ghimire; Wu,~S.; Aivazian,~G.; Ross,~J.~S.;
  Zhao,~B.; Yan,~J.; Mandrus,~D.~G.; Xiao,~D.; Yao,~W.; Xu,~X. \emph{Nature
  Nanotechnology} \textbf{2013}, \emph{8}, 634\relax
\mciteBstWouldAddEndPuncttrue \mciteSetBstMidEndSepPunct{\mcitedefaultmidpunct}
{\mcitedefaultendpunct}{\mcitedefaultseppunct}\relax \EndOfBibitem
\bibitem[Cao \latin{et~al.}(2012)Cao, Wang, Han, Ye, Zhu, Shi, Niu, Tan, Liu,
  and Feng]{Cao12}
Cao,~T.; Wang,~G.; Han,~W.; Ye,~H.; Zhu,~C.; Shi,~J.; Niu,~Q.; Tan,~P.; Liu,~E.
  W.~B.; Feng,~J. \emph{Nature Communications} \textbf{2012}, \emph{3},
  887\relax
\mciteBstWouldAddEndPuncttrue \mciteSetBstMidEndSepPunct{\mcitedefaultmidpunct}
{\mcitedefaultendpunct}{\mcitedefaultseppunct}\relax \EndOfBibitem
\bibitem[Mak \latin{et~al.}(2012)Mak, He, Shan, and Heinz]{Mak12}
Mak,~K.~F.; He,~K.; Shan,~J.; Heinz,~T.~F. \emph{Nature Nanotechnology}
  \textbf{2012}, \emph{7}, 494\relax
\mciteBstWouldAddEndPuncttrue \mciteSetBstMidEndSepPunct{\mcitedefaultmidpunct}
{\mcitedefaultendpunct}{\mcitedefaultseppunct}\relax \EndOfBibitem
\bibitem[Xu \latin{et~al.}(2014)Xu, Yao, Xiao, and Heinz]{Xu14}
Xu,~X.; Yao,~W.; Xiao,~D.; Heinz,~T.~F. \emph{Nature Physics} \textbf{2014},
  \emph{10}, 343\relax
\mciteBstWouldAddEndPuncttrue \mciteSetBstMidEndSepPunct{\mcitedefaultmidpunct}
{\mcitedefaultendpunct}{\mcitedefaultseppunct}\relax \EndOfBibitem
\bibitem[Suzuki \latin{et~al.}(2014)Suzuki, Sakano, Zhang, Akashi, Morikawa,
  Harasawa, Yaji, Kuroda, Miyamoto, Okuda, Ishizaka, Arita, and
  Iwasa]{Suzuki14}
Suzuki,~R.; Sakano,~M.; Zhang,~Y.~J.; Akashi,~R.; Morikawa,~D.; Harasawa,~A.;
  Yaji,~K.; Kuroda,~K.; Miyamoto,~K.; Okuda,~T.; Ishizaka,~K.; Arita,~R.;
  Iwasa,~Y. \emph{Nature Nanotechnology} \textbf{2014}, \emph{9}, 611\relax
\mciteBstWouldAddEndPuncttrue \mciteSetBstMidEndSepPunct{\mcitedefaultmidpunct}
{\mcitedefaultendpunct}{\mcitedefaultseppunct}\relax \EndOfBibitem
\bibitem[Behnia(2012)]{Behnia2012}
Behnia,~K. \emph{Nature Nanotechnology} \textbf{2012}, \emph{7}, 488\relax \mciteBstWouldAddEndPuncttrue
\mciteSetBstMidEndSepPunct{\mcitedefaultmidpunct} {\mcitedefaultendpunct}{\mcitedefaultseppunct}\relax \EndOfBibitem
\bibitem[Nebel(2013)]{Neber12}
Nebel,~C.~E. \emph{Nature Materials} \textbf{2013}, \emph{12}, 690\relax \mciteBstWouldAddEndPuncttrue
\mciteSetBstMidEndSepPunct{\mcitedefaultmidpunct} {\mcitedefaultendpunct}{\mcitedefaultseppunct}\relax \EndOfBibitem
\bibitem[{Wu} \latin{et~al.}(2013){Wu}, {Ross}, {Liu}, {Aivazian}, {Jones},
  {Fei}, {Zhu}, {Xiao}, {Yao}, {Cobden}, and {Xu}]{Wu13}
{Wu},~S.; {Ross},~J.~S.; {Liu},~G.-B.; {Aivazian},~G.; {Jones},~A.; {Fei},~Z.;
  {Zhu},~W.; {Xiao},~D.; {Yao},~W.; {Cobden},~D.; {Xu},~X. \emph{Nature
  Physics} \textbf{2013}, \emph{9}, 149--153\relax
\mciteBstWouldAddEndPuncttrue \mciteSetBstMidEndSepPunct{\mcitedefaultmidpunct}
{\mcitedefaultendpunct}{\mcitedefaultseppunct}\relax \EndOfBibitem
\bibitem[Mak \latin{et~al.}(2014)Mak, McGill, Park, and McEuen]{Mak14}
Mak,~K.~F.; McGill,~K.~L.; Park,~J.; McEuen,~P.~L. \emph{Science}
  \textbf{2014}, \emph{344}, 1489--1492\relax
\mciteBstWouldAddEndPuncttrue \mciteSetBstMidEndSepPunct{\mcitedefaultmidpunct}
{\mcitedefaultendpunct}{\mcitedefaultseppunct}\relax \EndOfBibitem
\bibitem[{Aivazian} \latin{et~al.}(2015){Aivazian}, {Gong}, {Jones}, {Chu},
  {Yan}, {Mandrus}, {Zhang}, {Cobden}, {Yao}, and {Xu}]{Aivazian2015}
{Aivazian},~G.; {Gong},~Z.; {Jones},~A.~M.; {Chu},~R.-L.; {Yan},~J.;
  {Mandrus},~D.~G.; {Zhang},~C.; {Cobden},~D.; {Yao},~W.; {Xu},~X. \emph{Nature
  Physics} \textbf{2015}, \emph{11}, 148\relax
\mciteBstWouldAddEndPuncttrue \mciteSetBstMidEndSepPunct{\mcitedefaultmidpunct}
{\mcitedefaultendpunct}{\mcitedefaultseppunct}\relax \EndOfBibitem
\bibitem[MacNeill \latin{et~al.}(2015)MacNeill, Heikes, Mak, Anderson,
  Korm\'anyos, Z\'olyomi, Park, and Ralph]{MacNeill2015}
MacNeill,~D.; Heikes,~C.; Mak,~K.~F.; Anderson,~Z.; Korm\'anyos,~A.;
  Z\'olyomi,~V.; Park,~J.; Ralph,~D.~C. \emph{Phys. Rev. Lett.} \textbf{2015},
  \emph{114}, 037401\relax
\mciteBstWouldAddEndPuncttrue \mciteSetBstMidEndSepPunct{\mcitedefaultmidpunct}
{\mcitedefaultendpunct}{\mcitedefaultseppunct}\relax \EndOfBibitem
\bibitem[{Srivastava} \latin{et~al.}(2015){Srivastava}, {Sidler}, {Allain},
  {Lembke}, {Kis}, and {Imamoglu}]{Srivastava15}
{Srivastava},~A.; {Sidler},~M.; {Allain},~A.~V.; {Lembke},~D.~S.; {Kis},~A.;
  {Imamoglu},~A. \emph{Nature Physics} \textbf{2015}, \emph{11}, 141\relax
\mciteBstWouldAddEndPuncttrue \mciteSetBstMidEndSepPunct{\mcitedefaultmidpunct}
{\mcitedefaultendpunct}{\mcitedefaultseppunct}\relax \EndOfBibitem
\bibitem[Li \latin{et~al.}(2014)Li, Ludwig, Low, Chernikov, Cui, Arefe, Kim,
  van~der Zande, Rigosi, Hill, Kim, Hone, Li, Smirnov, and Heinz]{Li14}
Li,~Y.; Ludwig,~J.; Low,~T.; Chernikov,~A.; Cui,~X.; Arefe,~G.; Kim,~Y.~D.;
  van~der Zande,~A.~M.; Rigosi,~A.; Hill,~H.~M.; Kim,~S.~H.; Hone,~J.; Li,~Z.;
  Smirnov,~D.; Heinz,~T.~F. \emph{Phys. Rev. Lett.} \textbf{2014}, \emph{113},
  266804\relax
\mciteBstWouldAddEndPuncttrue \mciteSetBstMidEndSepPunct{\mcitedefaultmidpunct}
{\mcitedefaultendpunct}{\mcitedefaultseppunct}\relax \EndOfBibitem
\bibitem[El-Mahalawy and Evans(1977)El-Mahalawy, and Evans]{Evans77}
El-Mahalawy,~S.~H.; Evans,~B.~L. \emph{Phys. Status Solidi B} \textbf{1977},
  \emph{79}, 713\relax
\mciteBstWouldAddEndPuncttrue \mciteSetBstMidEndSepPunct{\mcitedefaultmidpunct}
{\mcitedefaultendpunct}{\mcitedefaultseppunct}\relax \EndOfBibitem
\bibitem[Zeng \latin{et~al.}(2013)Zeng, Liu, Dai, Yan, Zhu, He, Xie, Xu, Chen,
  Yao, and Cui]{Zeng13}
Zeng,~H.; Liu,~G.-B.; Dai,~J.; Yan,~Y.; Zhu,~B.; He,~R.; Xie,~L.; Xu,~S.;
  Chen,~X.; Yao,~W.; Cui,~X. \emph{Scientific Reports} \textbf{2013}, \emph{3},
  1608\relax
\mciteBstWouldAddEndPuncttrue \mciteSetBstMidEndSepPunct{\mcitedefaultmidpunct}
{\mcitedefaultendpunct}{\mcitedefaultseppunct}\relax \EndOfBibitem
\bibitem[Zhao \latin{et~al.}(2013)Zhao, Ghorannevis, Chu, Toh, Kloc, Tan, and
  Eda]{Zhao13}
Zhao,~W.; Ghorannevis,~Z.; Chu,~L.; Toh,~M.; Kloc,~C.; Tan,~P.-H.; Eda,~G.
  \emph{ACS Nano} \textbf{2013}, \emph{7}, 791--797\relax
\mciteBstWouldAddEndPuncttrue \mciteSetBstMidEndSepPunct{\mcitedefaultmidpunct}
{\mcitedefaultendpunct}{\mcitedefaultseppunct}\relax \EndOfBibitem
\bibitem[Wang \latin{et~al.}(2014)Wang, Bouet, Lagarde, Vidal, Balocchi, Amand,
  Marie, and Urbaszek]{Urbaszek14}
Wang,~G.; Bouet,~L.; Lagarde,~D.; Vidal,~M.; Balocchi,~A.; Amand,~T.;
  Marie,~X.; Urbaszek,~B. \emph{Phys. Rev. B} \textbf{2014}, \emph{90},
  075413\relax
\mciteBstWouldAddEndPuncttrue \mciteSetBstMidEndSepPunct{\mcitedefaultmidpunct}
{\mcitedefaultendpunct}{\mcitedefaultseppunct}\relax \EndOfBibitem
\bibitem[Beal and Liang(1976)Beal, and Liang]{Beal76}
Beal,~A.~R.; Liang,~W.~Y. \emph{Journal of Physics C: Solid State Physics}
  \textbf{1976}, \emph{9}, 2459\relax
\mciteBstWouldAddEndPuncttrue \mciteSetBstMidEndSepPunct{\mcitedefaultmidpunct}
{\mcitedefaultendpunct}{\mcitedefaultseppunct}\relax \EndOfBibitem
\bibitem[{Mak} \latin{et~al.}(2013){Mak}, {He}, {Lee}, {Lee}, {Hone}, {Heinz},
  and {Shan}]{Mak13}
{Mak},~K.~F.; {He},~K.; {Lee},~C.; {Lee},~G.~H.; {Hone},~J.; {Heinz},~T.~F.;
  {Shan},~J. \emph{Nature Materials} \textbf{2013}, \emph{12}, 207--211\relax
\mciteBstWouldAddEndPuncttrue \mciteSetBstMidEndSepPunct{\mcitedefaultmidpunct}
{\mcitedefaultendpunct}{\mcitedefaultseppunct}\relax \EndOfBibitem
\bibitem[Mitioglu \latin{et~al.}(2013)Mitioglu, Plochocka, Jadczak, Escoffier,
  Rikken, Kulyuk, and Maude]{Mitioglu13}
Mitioglu,~A.~A.; Plochocka,~P.; Jadczak,~J.~N.; Escoffier,~W.; Rikken,~G. L.
  J.~A.; Kulyuk,~L.; Maude,~D.~K. \emph{Phys. Rev. B} \textbf{2013}, \emph{88},
  245403\relax
\mciteBstWouldAddEndPuncttrue \mciteSetBstMidEndSepPunct{\mcitedefaultmidpunct}
{\mcitedefaultendpunct}{\mcitedefaultseppunct}\relax \EndOfBibitem
\bibitem[Zhu \latin{et~al.}(2014)Zhu, Zhang, Glazov, Urbaszek, Amand, Ji, Liu,
  and Marie]{Zhu14}
Zhu,~C.~R.; Zhang,~K.; Glazov,~M.; Urbaszek,~B.; Amand,~T.; Ji,~Z.~W.;
  Liu,~B.~L.; Marie,~X. \emph{Phys. Rev. B} \textbf{2014}, \emph{90},
  161302\relax
\mciteBstWouldAddEndPuncttrue \mciteSetBstMidEndSepPunct{\mcitedefaultmidpunct}
{\mcitedefaultendpunct}{\mcitedefaultseppunct}\relax \EndOfBibitem
\bibitem[Rose \latin{et~al.}(2013)Rose, Goerbig, and Pi\'echon]{Rose2013}
Rose,~F.; Goerbig,~M.~O.; Pi\'echon,~F. \emph{Phys. Rev. B} \textbf{2013},
  \emph{88}, 125438\relax
\mciteBstWouldAddEndPuncttrue \mciteSetBstMidEndSepPunct{\mcitedefaultmidpunct}
{\mcitedefaultendpunct}{\mcitedefaultseppunct}\relax \EndOfBibitem
\bibitem[He \latin{et~al.}(2014)He, Kumar, Zhao, Wang, Mak, Zhao, and
  Shan]{He2014}
He,~K.; Kumar,~N.; Zhao,~L.; Wang,~Z.; Mak,~K.~F.; Zhao,~H.; Shan,~J.
  \emph{Phys. Rev. Lett.} \textbf{2014}, \emph{113}, 026803\relax
\mciteBstWouldAddEndPuncttrue \mciteSetBstMidEndSepPunct{\mcitedefaultmidpunct}
{\mcitedefaultendpunct}{\mcitedefaultseppunct}\relax \EndOfBibitem
\bibitem[Goto \latin{et~al.}(2000)Goto, Kato, Uchida, and Miura]{Goto2000}
Goto,~T.; Kato,~Y.; Uchida,~K.; Miura,~N. \emph{Journal of Physics: Condensed
  Matter} \textbf{2000}, \emph{12}, 6719\relax
\mciteBstWouldAddEndPuncttrue \mciteSetBstMidEndSepPunct{\mcitedefaultmidpunct}
{\mcitedefaultendpunct}{\mcitedefaultseppunct}\relax \EndOfBibitem
\end{mcitethebibliography}

\providecommand{\latin}[1]{#1} \providecommand*\mcitethebibliography{\thebibliography} \csname
@ifundefined\endcsname{endmcitethebibliography}
  {\let\endmcitethebibliography\endthebibliography}{}

\end{document}